\documentclass[11pt]{article}
\usepackage{moriond,epsfig}

\def\Journal#1#2#3#4{{#1} {\bf #2}, #3 (#4)}

\def\PLB{{\em Phys. Lett.}  B}
\def\PRL{\em Phys. Rev. Lett.}
\def\PRD{{\em Phys. Rev.} D}

\def\be{\begin{equation}}
\def\ee{\end{equation}}
\def\bea{\begin{eqnarray}}
\def\eea{\end{eqnarray}}

\begin{document}
\vspace*{4cm}
\title{RECENT B PHYSICS RESULTS FROM THE TEVATRON}

\author{SATYAJIT BEHARI \\ (for the CDF and D\O\ Collaborations) \vspace*{8pt}}

\address{Department of Physics, The Johns Hopkins University,\\
       3400 N. Charles Street, Baltimore, MD 21218, USA}

%%%%%%%%%%%%%%% A b s t r a c t %%%%%%%%%%%%%%%%%%%
\maketitle\abstracts{
We review recent $B$ physics results from the CDF and D\O\ 
experiments in $p\bar{p}$ collisions at $\sqrt{s}$ = 1.96 TeV. 
Using a data sample of 1.4-6.0 fb$^{-1}$ collected by the 
CDF II detector we present searches for New Physics in $B_{s}$
sector and some competitive results with $B$-factories in the
$B$~/charm sector. In the first category we report the BR in
$B_s \to J/\psi f_0(980)$ decays and the time-integrated mixing 
probability ($\bar{\chi}$) of $B$ mesons. In the second 
category BR and $A_{CP}$ in doubly Cabibbo-suppressed 
$B^{\pm} \to D^0 h^{\pm}$ decays and time-integrated CP 
violation in $D^0 \to h^+ h^-$ are presented.
}

%%%%%%%%%%%%%%%%%%%%%%%%%%%%%%%%%%%%%%%%%%%%%%%%%%%%%%%%%%%%%
\section{Search for New Physics in $B_s$ decays}
\label{sec:NpInBs}
%%%%%%%%%%%%%%%%%%%%%%%%%%%%%%%%%%%%%%%%%%%%%%%%%%%%%%%%%%%%%
Our current understanding of $B_s$ physics, within the Standard
Model (SM) and its sensitivity to New Physics (NP), is derived
exclusively from the large Tevatron Run II data samples. One of 
the most interesting topics in this area is the measurement of
$B_s$ mixing phase, $\beta_{s}$. This phase is expected 
to be tiny in the Standard Model (SM):
\begin{eqnarray}
\beta_{s}^{SM}=arg(-V_{ts}V^*_{tb}/V_{cs}V^*_{cb})\approx 0.02,
\end{eqnarray}
and it is unconstrained by the 2006 measurements of the $B_s$ 
mixing frequency. Presence of NP can lead to large values of this
phase which is not excluded experimentally yet.

The $\beta_{s}^{SM}$ phase can be accessed through measuring the
time evolution of flavor-tagged $B_s \to J/\psi \phi$ decays, or 
inclusively by measuring the anomalous mixing rate difference, 
$A^b_{sl}$, between $B_s$ and $\bar{B_{s}}$. Both these methods 
are pursued at CDF and D\O\ .

Using 5.2 fb$^{-1}$ and 6.1 fb$^{-1}$ data samples, respectively, 
CDF and D\O\ have performed detailed angular analyses of the 
$B_s \to J/\psi \phi$ decays to disentangle their CP-even and 
CP-odd components. Their initial results indicate departure from 
SM by 0.8$\sigma$~\cite{CDF_betas} and 1.1$\sigma$~\cite{D0_betas}, 
calling for more scrutiny through independent measurements.

%%%%%%%%%%%%%%%%%%%%%%%%%%%%%%%%%%%%%%%%%%%%%%%%%%%%%%%%%%%%%%%%%%%%%%
\subsection{Measurement of $BR(B_s \to J/\psi f_0(980))$}
\label{subsec:f0}
%%%%%%%%%%%%%%%%%%%%%%%%%%%%%%%%%%%%%%%%%%%%%%%%%%%%%%%%%%%%%%%%%%%%%%
A simpler way to measure $\beta_s$ is through the study of 
$B_s \to J/\psi f_0(980), f_0(980) \to \pi^+ \pi^-$ decays. This 
is a pure CP-odd decay which can provide a clean measurement of
$\beta_s$ without a need for angular analysis. As a first step
towards this CDF searched for this suppressed decay mode using
3.8 fb$^{-1}$ data~\cite{CDF_f0}, collected using a di-muon trigger.

The search for $B_s \to J/\psi f_0(980)$ decays proceeds through
an initial loose selection of $\mu\mu\pi\pi$ candidates, followed 
by a Neural Network discrimination, based on kinematic variables, 
track and vertex displacement and isolation, for efficient
background suppression. An identical selection is used for the 
$B_s \to J/\psi \phi$ reference mode. A simultanous 
log-likelihood fit to the signal and normalization modes yields 
502$\pm$37(stat.)$\pm$18(syst.) $B_s \to J/\psi f_0(980)$ 
and 2302$\pm$499(stat.)$\pm$49(syst.) $B_s \to J/\psi \phi$ candidates.

\begin{figure}[h]
\centerline{\includegraphics[width=0.50\textwidth]{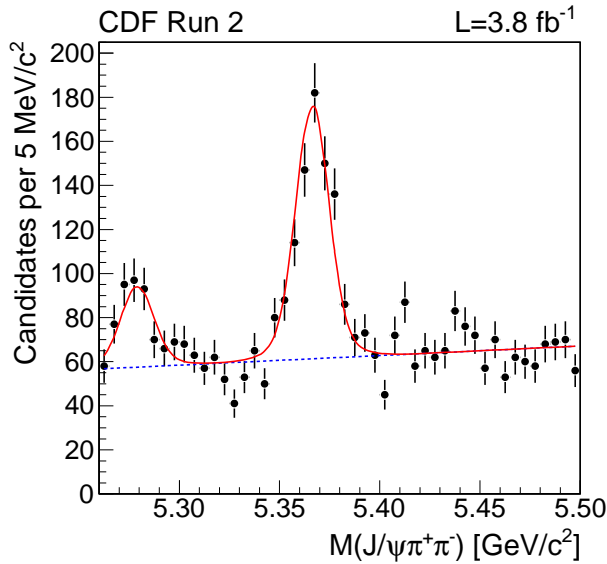}}
\caption{The invariant mass of $J\psi\pi\pi$ candidate events,
       measured by CDF with 3.8 fb$^{-1}$ of data.}
\label{fig:f0}
\end{figure}

Fig.~\ref{fig:f0} shows the invariant mass distribution of the 
$J/\psi\pi\pi$ candidate events.
In addition to the $B_s$ peak with a significance of observation 
of 17.9$\sigma$, also seen is the $B^0 \to J/\psi\rho^0$ peak.
The ratio between $BR(B_s \to J/\psi f_0(980), f_0(980)\to\pi\pi)$ 
and $BR(B_s \to J/\psi\phi, \phi \to KK)$ candidates, 
$R_{f_{0}/\phi}$, is 0.257 $\pm$ 0.020 (stat.) $\pm$ 0.014(syst.), 
resulting in a measurement of the branching ratio
\begin{eqnarray}
BR(B_s\to J/\psi f_0(980), f_0(980)\to\pi\pi)=1.63\pm0.12(stat.)\pm
 0.09(syst.)\pm0.50(PDG)\times 10^{-4}.
\end{eqnarray}
This is the most precise result obtained to date and confirms 
earlier resuts from Belle~\cite{Belle_f0} and LHC-b~\cite{LHCb_f0}.

%%%%%%%%%%%%%%%%%%%%%%%%%%%%%%%%%%%%%%%%%%%%%%%%%%%%%%%%%%%%%%%%%%%%%%%%%%%%%%%
\subsection{Time-integrated mixing probability ($\bar{\chi}$) of 
          $B$ mesons}
\label{subsec:chibar}
%%%%%%%%%%%%%%%%%%%%%%%%%%%%%%%%%%%%%%%%%%%%%%%%%%%%%%%%%%%%%%%%%%%%%%%%%%%%%%%
One of the most exciting results from Tevatron in recent times
is the dimuon charge asymmetry measurement by the D\O\ 
Collaboration using muon pairs produced in semileptonic decays
of $b$ hadrons~\cite{D0_absl}. The asymmetry $A^b_{sl}$ is 
defined as
\begin{eqnarray}
A^b_{sl}= \frac{N^{++}_b-N^{--}_b}{N^{++}_b+N^{--}_b},
\end{eqnarray}
where $N^{++}_b$ and $N^{--}_b$ are the numbers of same sign 
dimuon events produced due to the $b$ hadrons decaying 
semileptonically, one before and the other after mixing. The 
quantity $A^b_{sl}$ is expected to be approximately zero within 
the SM ($\approx$ a few 10$^{-4}$), if mixing rates 
($B\to \bar{B}$) and ($\bar{B}\to B$) are equal.
Using a 6.1 fb$^{-1}$ Run II data sample D\O\ measured
$A^b_{sl}=(-0.957\pm 0.251(stat.)\pm 0.146(syst.))\%$, which 
differs from the SM prediction~\cite{Nierste_absl} of 
$A^b_{sl}(SM)=(-0.023^{+0.005}_{-0.006})\%$ at about 3.2$\sigma$, 
indicating an anomalously large $B_s$ mixing phase.

The CDF Collaboration is pursuing an alternate path for
independent verification of the interesting D\O\ result. The same
sign (SS) and opposite sign (OS) muon impact parameter (IP)
distributions are fitted separately to disentangle the long-lived
di-muon component that originates from $B$ decays. The IP fitting 
method is a robust technique demonstrated well in the correlated 
$B\bar{B}$ cross-section measurement~\cite{CDF_bbxsec}.

\begin{figure}[h]
\centerline{\includegraphics[width=0.40\textwidth]{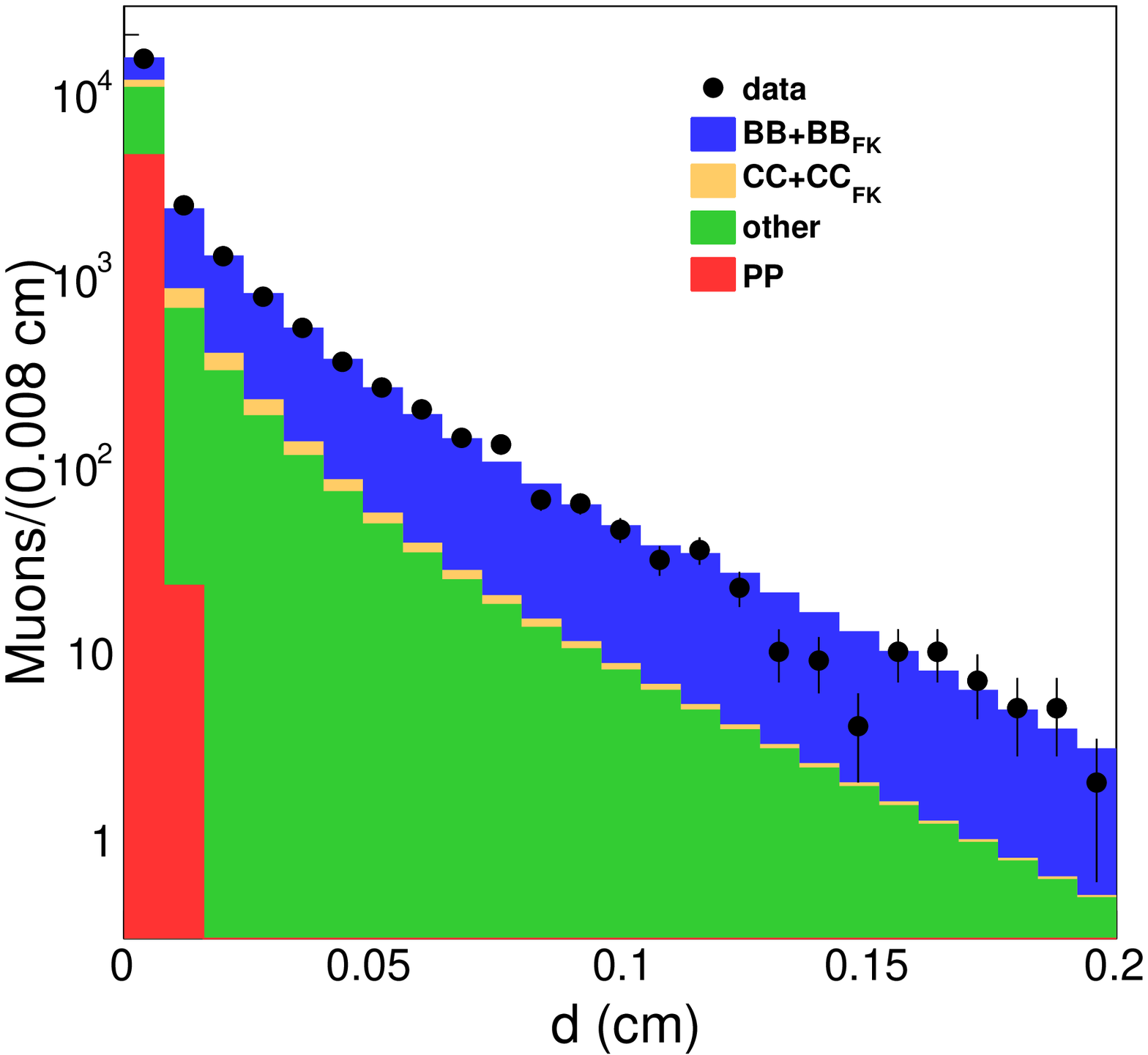}
            \includegraphics[width=0.40\textwidth]{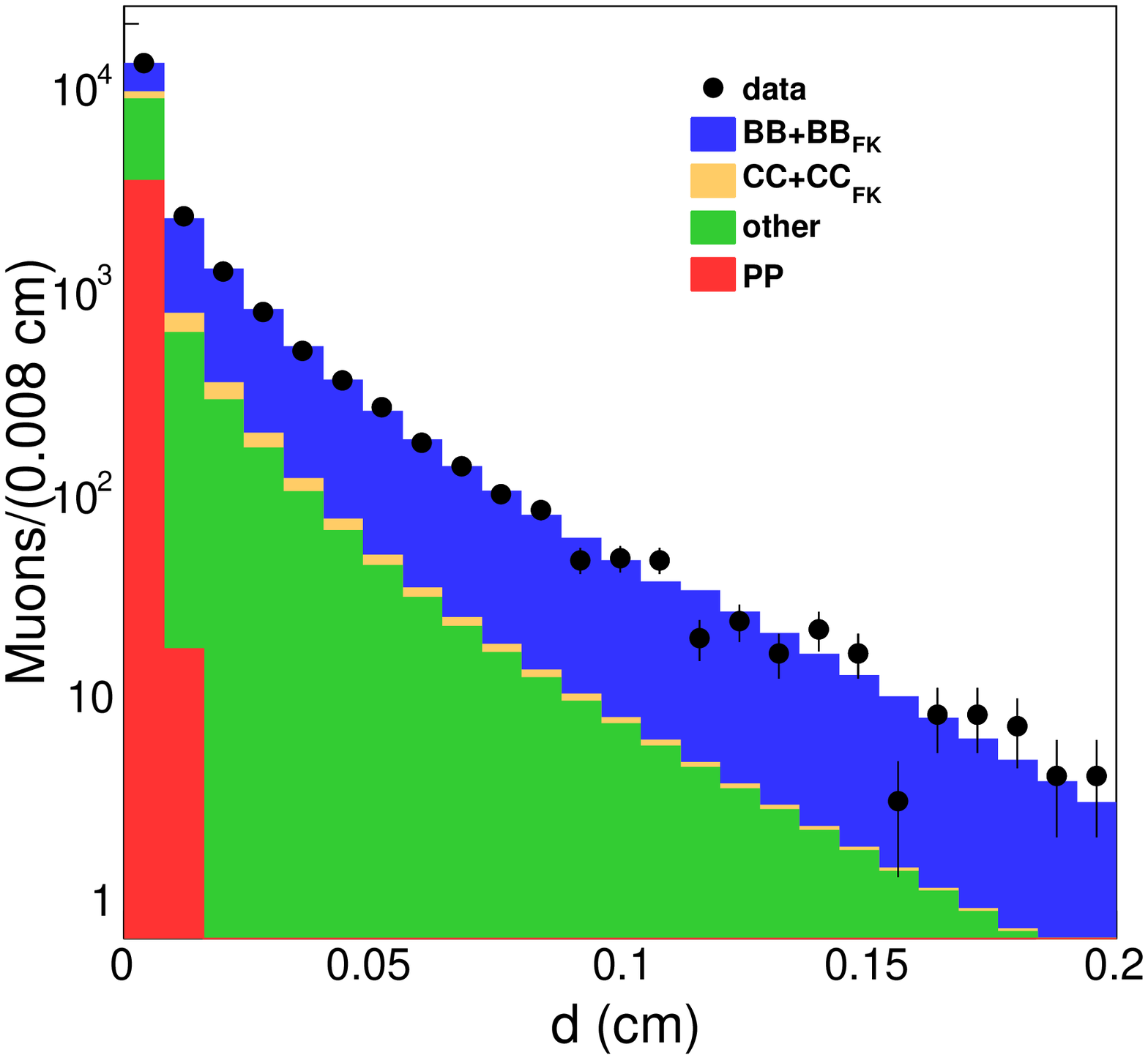}}
\caption{The muon impact parameter (IP) distributions are shown
       for same sign (a) $\mu^+\mu^+$ and (b) $\mu^-\mu^-$ pairs
       with IP fit results for $b$, $c$, prompt and other sources.}
\label{fig:chib}
\end{figure}

As a first step towards an $A^b_{sl}$ measurement, CDF further
puts to test the IP fitting method and measures the 
time-integrated mixing probability, $\bar{\chi}$~\cite{CDF_chib}, 
defined as
\begin{eqnarray}
\bar{\chi}=\frac{\Gamma(B^0_{d,s}\to\bar{B}^0_{d,s}\to
 l^+X)}{\Gamma(B_{all}\to l^{\pm}X)}= f_d\cdot\chi_d + f_s\cdot\chi_s
\end{eqnarray}
where $f_{d,s}$ are production fractions and $\chi_{d,s}$ are
mixing probabilities of $B_d$ and $B_s$ mesons. The number of
OS and SS muon pairs is measured and $\bar{\chi}$ is extracted 
from the ratio
$R=[N(\mu^+\mu^+)+N(\mu^-\mu^-)]/N(\mu^+\mu^-)$. Fig.~\ref{fig:chib}
shows the muon IP distributions for same sign dimuon pairs, with
IP fit results for $b$, $c$, prompt and other sources.
%
%A 2004 measurement from CDF~\cite{CDF_old_chib}, showed a 
%discrepancy with an earlier result from LEP.
%
Using a  1.4 fb$^{-1}$ data sample CDF measures an SS to OS ratio
of R = 0.467$\pm$0.011, leading to $\bar{\chi}$=0.126$\pm$0.008,
in very good agreement with the LEP result of 
$\bar{\chi}$=0.126$\pm$0.004. This validates the IP fitting method
and presents an encouraging prospect towards CDF's future $A_{sl}$ 
measurement.

%%%%%%%%%%%%%%%%%%%%%%%%%%%%%%%%%%%%%%%%%%%%%%%%%%%%%%%
\section{New Tevatron results in the $B$~/charm sector}
\label{sec:BCharm}
%%%%%%%%%%%%%%%%%%%%%%%%%%%%%%%%%%%%%%%%%%%%%%%%%%%%%%%
Over the past decade, the Belle and BaBar $B$-factories 
have shaped up our knowledge of the $B_{u,d}$ and charm physics.
With accumulation of large volume of Tevatron Run II data, the 
CDF and D\O\ experiments have caught up with, and in some cases 
surpassed, them in precision. In this section we present two new 
measurements from the CDF collaboration, competitive with the
$B$-factories.

%%%%%%%%%%%%%%%%%%%%%%%%%%%%%%%%%%%%%%%%%%%%%%%%%%%%%%%%%%%%%%%%%%%%%%%
\subsection{BR and $A_{CP}$ in $B^{\pm}\to D^0h^{\pm}$ decays}
\label{subsec:dcs}
%%%%%%%%%%%%%%%%%%%%%%%%%%%%%%%%%%%%%%%%%%%%%%%%%%%%%%%%%%%%%%%%%%%%%%%
The branching fractions and searches for CP asymmetries in 
$B^{\pm}\to D^0h^{\pm}$ decays allow for a theoretically clean 
measurement of $\gamma$, the least well constrained angle of the 
CKM matrix (known to 10-20$^{\circ}$ level). The proposed ADS 
method~\cite{ads} relies on interference between $B^{\pm}$ decay
modes proceeding through color allowed and suppressed modes
followed by $D^0$ decay via doubly Cabibbo-suppressed (DCS) and 
Cabbibo-favoured modes, respectively, which can lead to large 
$A_{CP}$.

The DCS fraction and asymmetry in $B\to D K$ decays are defined as:

\begin{eqnarray}
R_{ADS}(K)=\frac{BR(B^-\to[K^+\pi^-]_DK^-)+BR(B^+\to[K^-\pi^+]_DK^+)}
               {BR(B^-\to[K^-\pi^+]_DK^-)+BR(B^+\to[K^+\pi^-]_DK^+)}\\
A_{ADS}(K)=\frac{BR(B^-\to[K^+\pi^-]_DK^-)-BR(B^+\to[K^-\pi^+]_DK^+)}
               {BR(B^-\to[K^-\pi^+]_DK^-)+BR(B^+\to[K^+\pi^-]_DK^+)}\\
\end{eqnarray}

\begin{figure}[h]
\centerline{\includegraphics[width=0.40\textwidth]{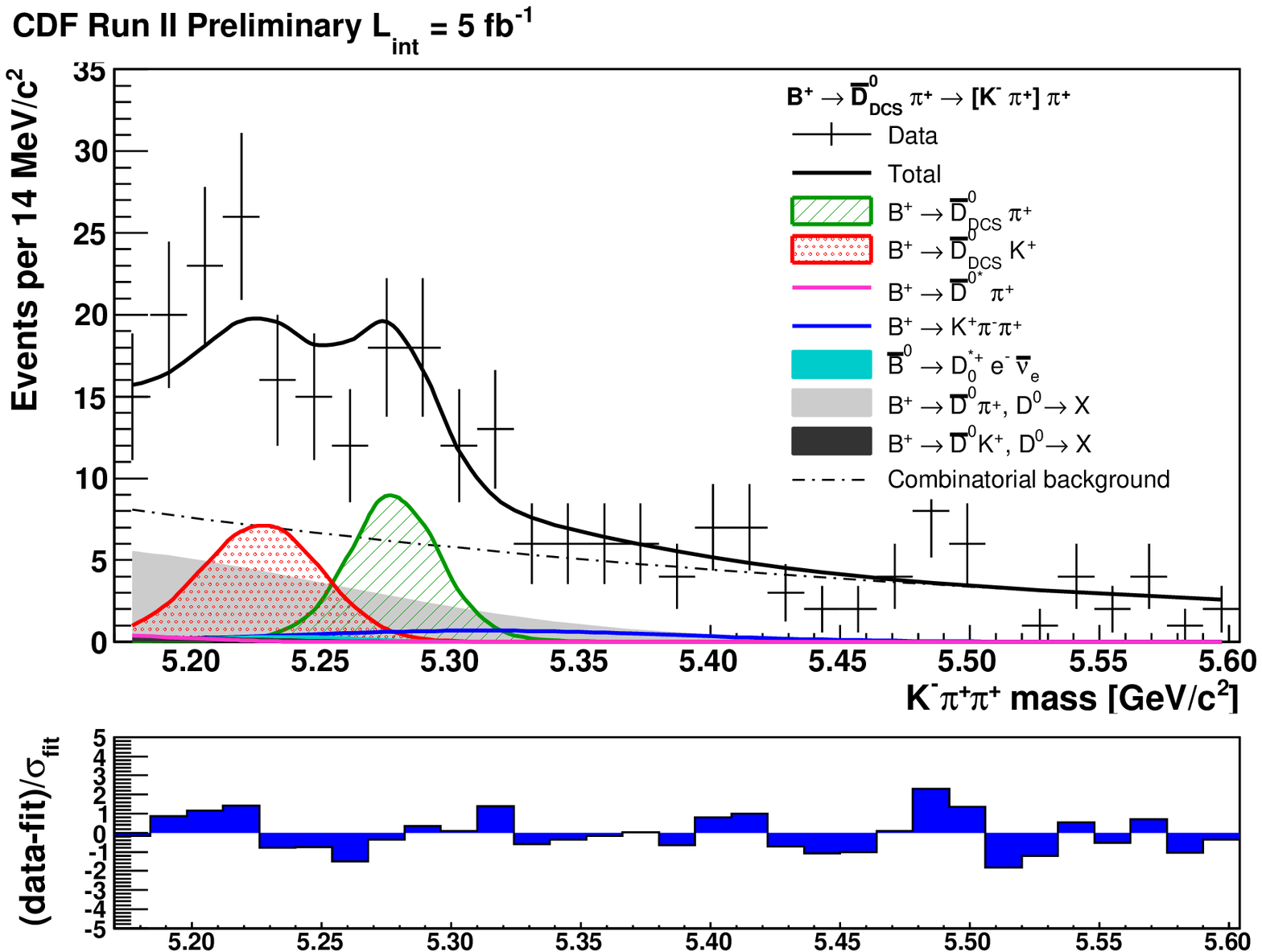}
            \includegraphics[width=0.40\textwidth]{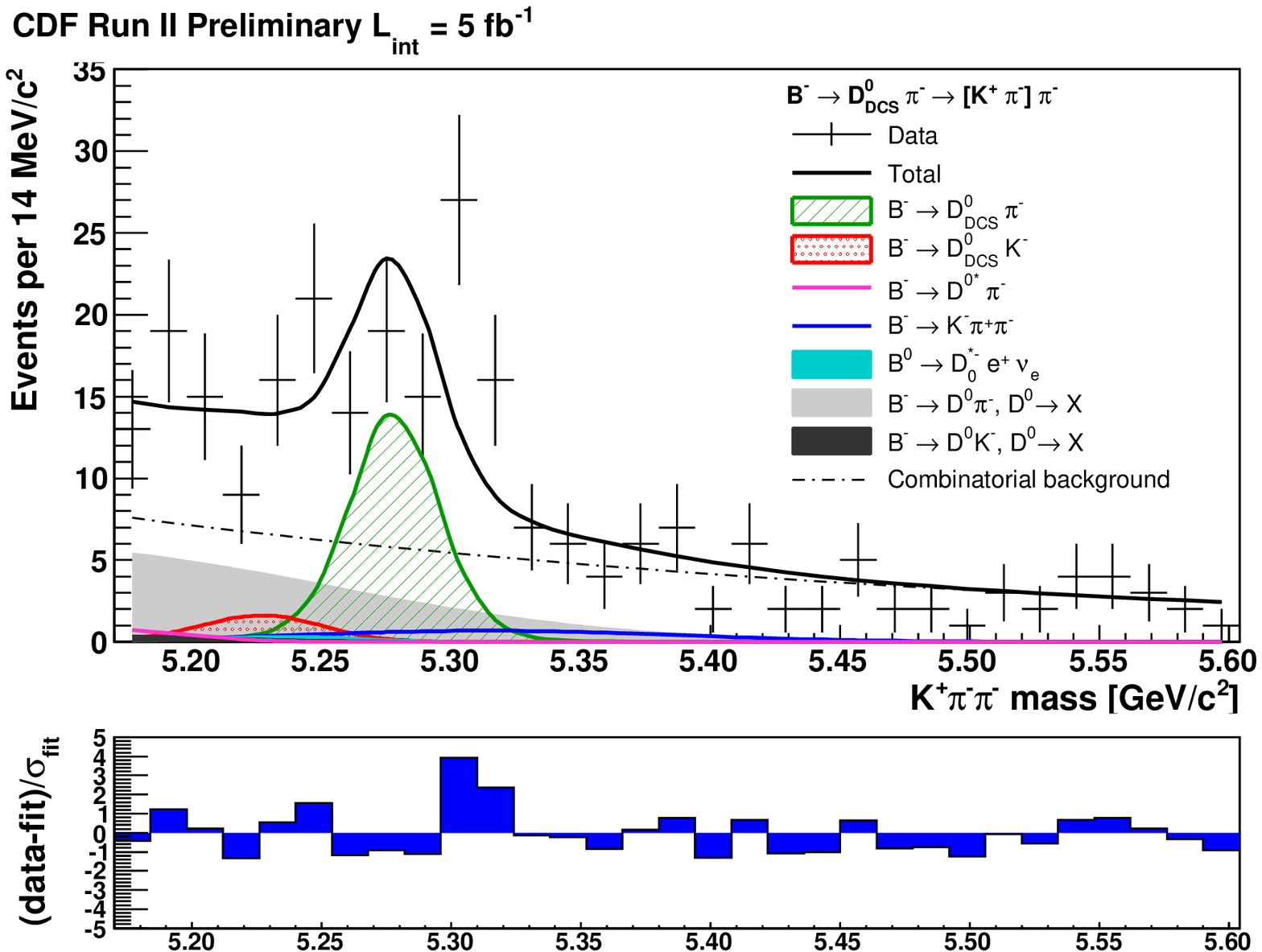}}
\caption{Invariant $K\pi\pi$ mass distributions for data and various 
background and signal contributions, separately for $B^+$ and 
$B^-$ decays.}
\label{fig:dcs}
\end{figure}

\begin{figure}[h]
\centerline{\includegraphics[width=0.40\textwidth]{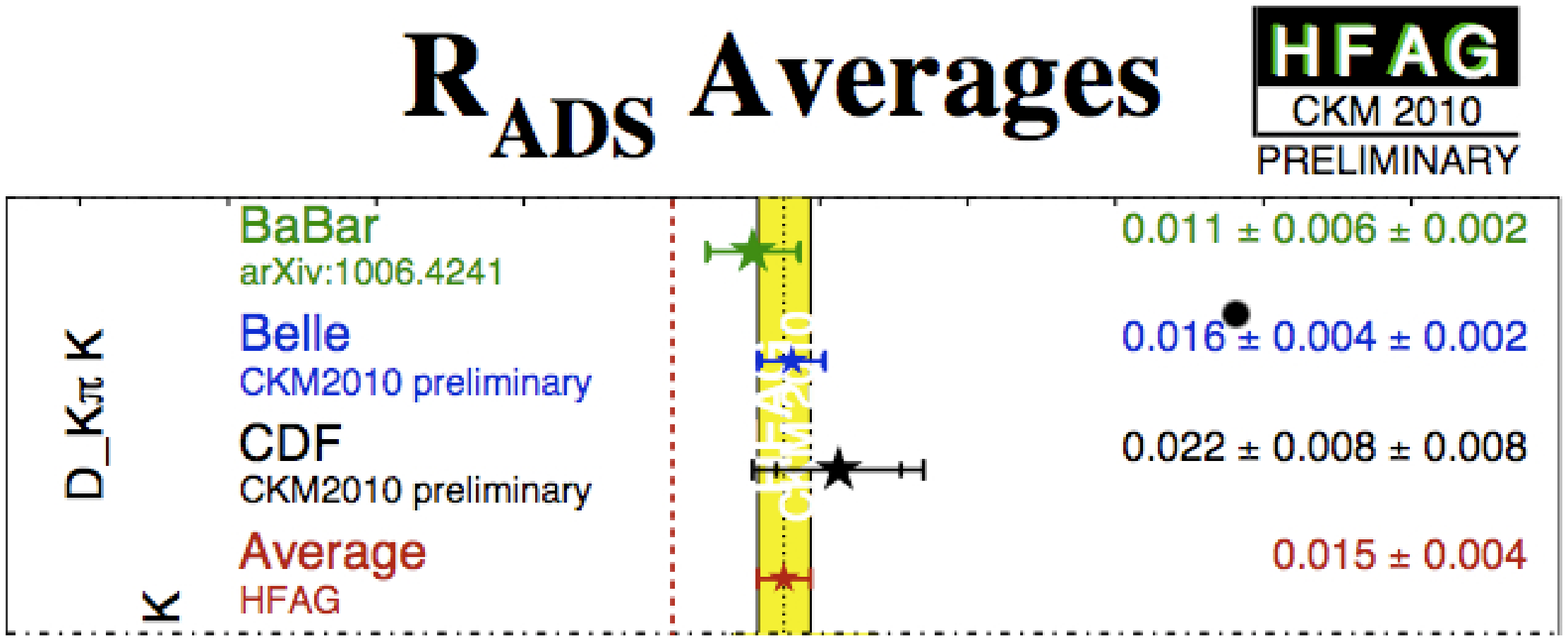}
            \includegraphics[width=0.40\textwidth]{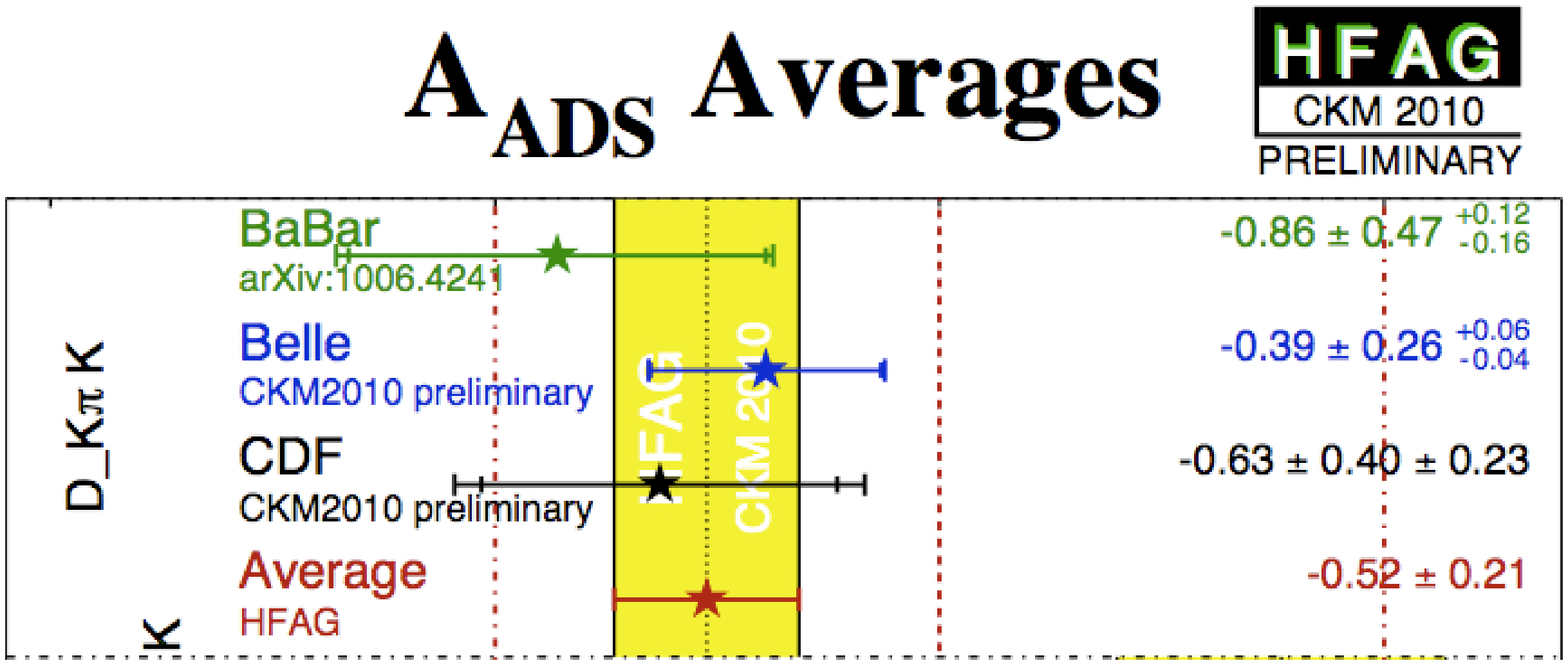}}
\caption{The DCS fraction and asymmetry results for the Kaon mode.}
\label{fig:ADS_results}
\end{figure}

Similar quantities for pions, $R_{ADS}(\pi)$ and $A_{ADS}(\pi)$,
can be defined. The experimental challenge is to suppress the
combinatorial and physics backgrounds when extracting the highly 
suppressed DCS signal. Using 5 fb$^{-1}$ of CDF Run II data, a 
likelihood fit combining mass and particle ID information is used 
to distinguish the signal ($D^0\pi$ and $D^0K$) modes from the 
background~\cite{CDF_ads}. Fig.~\ref{fig:dcs} shows the invariant 
$K\pi\pi$ mass distributions for DCS signal modes separately for 
$B^+$ and $B^-$ decays.
The yields for $\pi$ and $K$ modes are 73$\pm$16 and
34$\pm$14, respectively. The DCS fraction and asymmetry results 
for the Kaon mode are shown in Fig.~\ref{fig:ADS_results}, 
demonstrating good agreement with those from BaBar and Belle.
This is the first application of the ADS method at a hadron 
machine. CDF's new measurement of direct CP asymmetry for the 
DCS modes will be used in the future to extract $\gamma$.

%%%%%%%%%%%%%%%%%%%%%%%%%%%%%%%%%%%%%%%%%%%%%%%%%%%%%%%%%%%%%%%%%%%%%%%%%%%%%
\subsection{Time-integrated $A_{CP}$ in $D^0\to h^+h^-$ decays}
\label{subsec:dhh}
\begin{figure}[h]
\centerline{\includegraphics[width=0.40\textwidth]{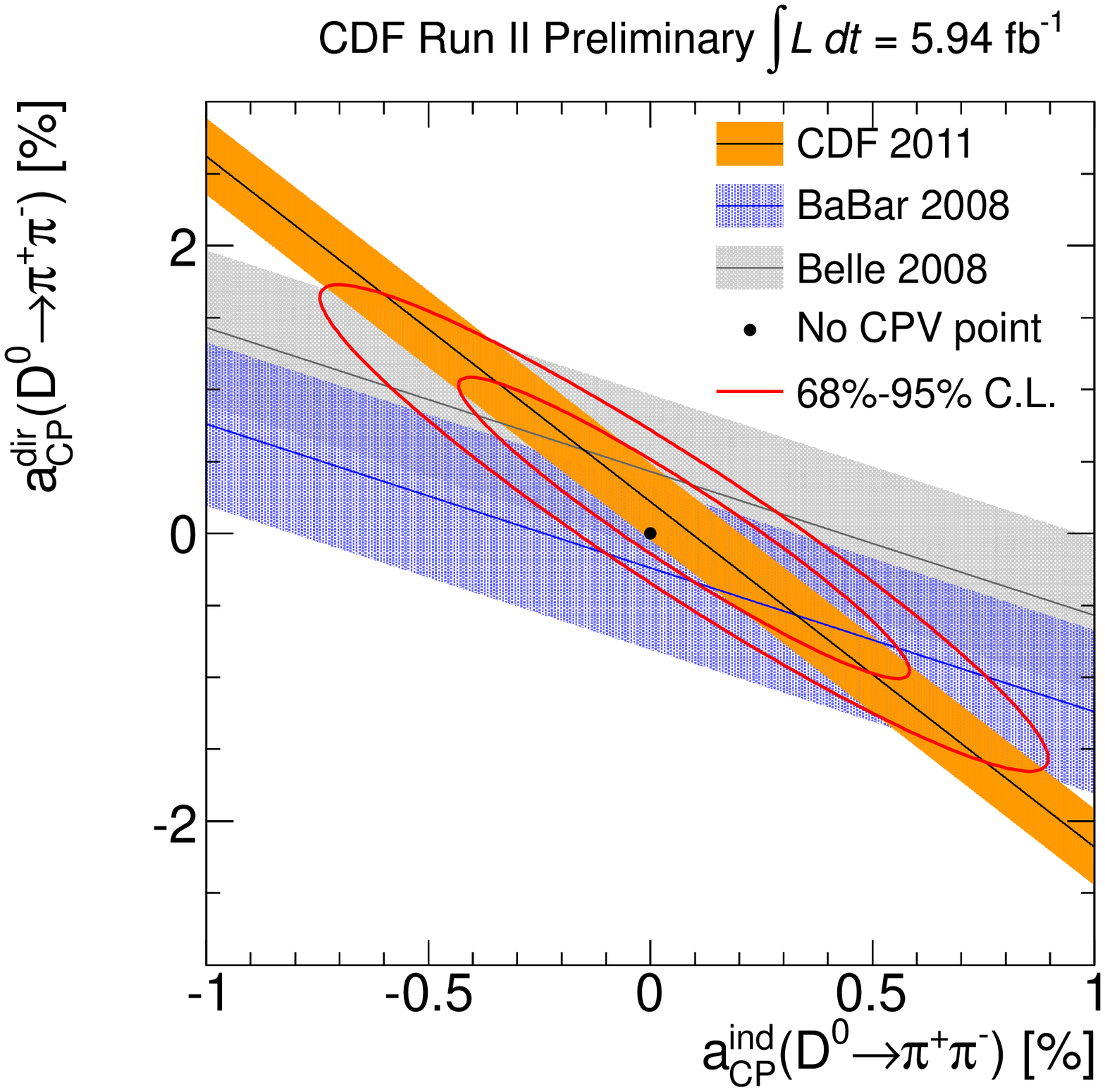}
            \includegraphics[width=0.40\textwidth]{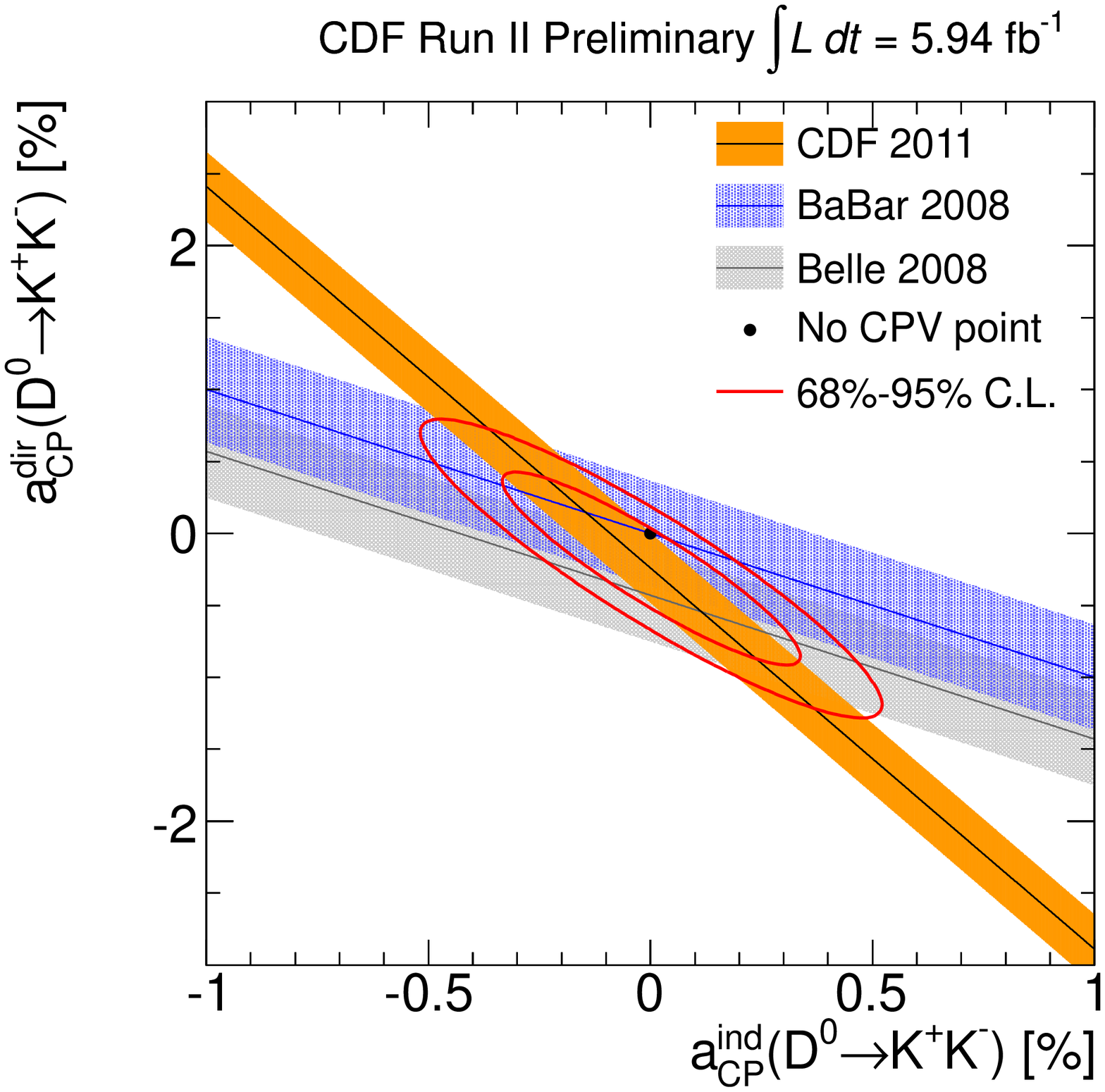}}
\caption{CP asymmetries in $D^0\to\pi\pi$ and $D^0\to KK$ decay
        modes and combination (contour) with the $B$-factory
        results.}
\label{fig:dhh_results}
\end{figure}

%%%%%%%%%%%%%%%%%%%%%%%%%%%%%%%%%%%%%%%%%%%%%%%%%%%%%%%%%%%%%%%%%%%%%%%%%%%%%
CP violation in the charm sector has been an area of great 
interest. Recent studies~\cite{grossman} have pointed out that, 
similar to $D^0$ oscillations, NP contributions could play a 
role in enhancing the size of CP violation in the charm sector.
Since in SM there is negligible penguin contribution to the
charm decays, an $A_{CP}$ larger than $\sim$0.1\% would be a
clear indication of NP. The relevant asymmetry is defined as
\begin{eqnarray}
A_{CP}(h^+h^-)=\frac{\Gamma(D^0\to h^+h^-)-\Gamma(\bar{D^0}\to h^+h^-)}
                   {\Gamma(D^0\to h^+h^-)+\Gamma(\bar{D^0}\to h^+h^-)}.
\end{eqnarray}
Using a 5.94 fb$^{-1}$ data sample of self-tagged 
$D^{*\pm}\to D^0\pi^{\pm}_s\to [h^+h^-]\pi^{\pm}_s$ decays CDF
extracts clean $D^0\to h^+h^-$ samples~\cite{CDF_dhh}. The 
asymmetry in $\pi\pi$ 
and $KK$ samples is measured and corrected for the instrumental 
asymmetry using $K\pi$ samples, with and without the $D^*$ tag.  
Fig.~\ref{fig:dhh_results} shows the CP asymmetries for
$D^0\to\pi\pi$ and $D^0\to KK$ decay modes and 68\%-95\% C.L. for
their combination with the $B$-factory results.
The resulting asymmetries are world's best:
\begin{eqnarray}
A_{CP}(D^0\rightarrow \pi^+\pi^-)=[+0.22\pm0.24(stat.)\pm0.11(syst.)]\%\\
A_{CP}(D^0\rightarrow K^+K^-)=[-0.24\pm0.22(stat.)\pm 0.10(syst.)]\%
\end{eqnarray}
and are compatible with Belle/BaBar 2008 results.

%%%%%%%%%%%%%%%%%%%%%%%
\section{Conclusion}
%%%%%%%%%%%%%%%%%%%%%%%
The Tevatron will continue to produce a steady flow of high 
luminosity data until Oct. 2011 when it is scheduled to shutdown 
permanently. Since mid-2000's the Tevatron experiments have
emerged as leaders in the field of Heavy Flavor physics and would
leave behind a rich legacy for current/future experiments through 
many landmark measurements. This article reviews a few of the 
interesting spring 2011 Heavy Flavor results from the Tevatron, 
which use a fraction of the accumulated data. With many more 
analyses in the pipeline and more data to be analyzed we look 
forward to challenging the SM predictions and constraining NP 
model parameters in the months and years to come.

%%%%%%%%%%%%%%%%%%%%%%%%%%%%%%%
\section*{Acknowledgments}
%%%%%%%%%%%%%%%%%%%%%%%%%%%%%%%
The author would like to thank his CDF and D\O\ collaborators for 
their help in preparing this contribution and the conference
organizers/secretariate for making the event an unique and 
thoroughly enjoyable experience.

%%%%%%%%%%%%%%%%%%%%%%%%%
\section*{References}
%%%%%%%%%%%%%%%%%%%%%%%%%

\end{document}